\documentclass[smallcondensed]{svjour3}    

\smartqed  

\usepackage{graphicx}
\usepackage{amssymb,amsmath}
\usepackage{natbib}
\usepackage{txfonts}
\usepackage{url}
\usepackage{pifont}
\usepackage{multirow}
\usepackage{hhline}

\def\G{\mathcal{G}}

\newcommand{\Msun}{M_\odot}
\newcommand{\Rsun}{R_\odot}

\newcommand{\Ms}{M_{\star}}
\newcommand{\Rs}{R_{\star}}

\def\Mearth{ M_\oplus}

\newcommand{\Mp}{M_{\rm p}}

\newcommand{\Mjup}{ M_{\rm Jup}}

\newcommand{\Os}{\Omega_{\star}}
\newcommand{\sss}{\sigma_{\star}}
\newcommand{\oss}{\overline{\sss}}
\newcommand{\Q}{\overline{Q'}}
\newcommand{\Qs}{\overline{Q_s'}}
\newcommand{\Ts}{T_\star}

\newcommand{\dd}{\mathrm{d}}

\usepackage[normalem]{ulem}
\usepackage{color}
\definecolor{blue}{RGB}{0,0,255}
\definecolor{red}{RGB}{255,0,0}
\definecolor{green}{RGB}{0,200,0}
\definecolor{black}{RGB}{0,0,0}

\begin{document}

\title{Effect of the rotation and tidal dissipation history of stars on the evolution of close-in planets}


\author{Emeline Bolmont         \and
        St\'ephane Mathis 
}


\institute{Emeline Bolmont \at
              NaXys, Department of Mathematics, University of Namur, 8 Rempart de la Vierge, 5000 Namur, Belgium \\
              Tel.: +32 81 72 49 19\\
              \email{emelinebolmont@unamur.be}
              \and St\'ephane Mathis \at
              Laboratoire AIM Paris-Saclay, CEA/DRF - CNRS - Univ. Paris Diderot - IRFU/SAp, Centre de Saclay, F-91191 Gif-sur-Yvette Cedex, France}        
\maketitle

\begin{abstract}
Since twenty years, a large population of close-in planets orbiting various classes of low-mass stars (from M-type to A-type stars) has been discovered. 
In such systems, the dissipation of the kinetic energy of tidal flows in the host star may modify its rotational evolution and shape the orbital architecture of the surrounding planetary system. 
In this context, recent observational and theoretical works demonstrated that the amplitude of this dissipation can vary over several orders of magnitude as a function of stellar mass, age and rotation. 
In addition, stellar spin-up occurring during the Pre-Main-Sequence (PMS) phase because of the contraction of stars and their spin-down because of the torque applied by magnetized stellar winds strongly impact angular momentum exchanges within star-planet systems. 
Therefore, it is now necessary to take into account the structural and rotational evolution of stars when studying the orbital evolution of close-in planets. 
At the same time, the presence of planets may modify the rotational dynamics of the host stars and as a consequence their evolution, magnetic activity and mixing.
In this work, we present the first study of the dynamics of close-in planets of various masses orbiting low-mass stars (from $0.6~\Msun$ to $1.2~\Msun$) where we compute the simultaneous evolution of the star's structure, rotation and tidal dissipation in its external convective enveloppe. 
We demonstrate that tidal friction due to the stellar dynamical tide, i.e. tidal inertial waves excited in the convection zone, can be larger by several orders of magnitude than the one of the equilibrium tide currently used in celestial mechanics, especially during the PMS phase. 
Moreover, because of this stronger tidal friction in the star, the orbital migration of the planet is now more pronounced and depends more on the stellar mass, rotation and age. This would very weakly affect the planets in the habitable zone because they are located at orbital distances such that stellar tide-induced migration happens on very long timescales. 
We also demonstrate that the rotational evolution of host stars is only weakly affected by the presence of planets except for massive companions.

\keywords{Planets and satellites: dynamical evolution and stability -- Planet-star interactions -- Planets and satellites: terrestrial planets -- Planets and satellites: gaseous planets -- stars: evolution -- stars: rotation}

\end{abstract}

\section{Introduction}

The discovery twenty years ago of the first hot Jupiter 51 Peg close to its host star \citep{MayorQueloz1995} opened the path to the detection and characterization of more than 1600 confirmed exoplanetary systems, which have a large diversity of host stars (from M to A type-stars), orbital architecture and planetary types \citep[e.g.][]{Perryman2011, Fabryckyetal2014}. 
Among them, a large population of systems are constituted by planets orbiting very close to their host stars as this is the case for example for hot Jupiters \citep[e.g.][]{MayorQueloz1995, Henry2000, Charbonneau2000} and other compact planetary systems \citep[e.g.][]{FangMargot2012, Fabryckyetal2014}. 
In those configurations, tidal dissipation inside the host stars may strongly affects surrounding planetary orbits \citep{Jackson2008, Husnooetal2012, Lai2012, Guillotetal2014} and the spin-orbit inclination \citep[e.g.][]{Winnetal2010, Albrechtetal2012} while the presence of a massive close-in planet should modify its rotational evolution \citep[e.g.][]{Pont2009, Lanza2010, Bolmont2012, McQuillan2013, Pop2014, Paz2015, Ferraz-Mello2015, Ceillieretal2016}.\\ 

At the same time, stellar structure and rotation strongly vary from the formation of their planetary systems during their Pre-Main-Sequence to their late stage of evolution \citep[e.g.][and references therein]{Maeder2009, Bouvier2008, Irwin2011, McQuillan2013, McQuillanetal2014}. 
This has strong consequences for the amplitude of tidal dissipation in their interiors \citep[e.g.][]{Zahn1966, ZahnBouchet1989, OgilvieLin2007, Mathis2015a, Mathis2015b}, their magnetic activity \citep[e.g.][]{Barnes2003, BarnesKim2010, Garcia2014} and their related winds \citep[e.g.][]{Skumanich1972, Kawaler1988}. 
In stellar convective regions, tidal dissipation is due to the action of the turbulent friction applied by convective eddies on the so-called equilibrium tide flow induced by the hydrostatic elongation along the line of centers because of the presence of the companion \citep{Zahn1966, RMZ2012} and tidal inertial waves driven by the Coriolis acceleration \citep{OgilvieLin2007, GL2009, Lai2012, AuclairDesrotouretal2015}. 
In stellar radiation zones, it is due to thermal diffusion and breaking mechanisms acting on gravito-inertial waves \citep{Zahn1975, BarkerOgilvie2010}. In the case of the convective enveloppe of low-mass stars recent works \citep{OgilvieLin2007,BO2009,Mathis2015a} demonstrated that tidal dissipation vary over several orders of magnitude with stellar mass, age and the related internal structure, and rotation. 
Simultaneously, because of their rotational evolution, the torques applied by stellar winds vary with stellar age \citep{Matt2015}.\\

Since tidal dissipation and stellar winds both strongly impact the dynamical evolution of their planetary systems along the evolution of their host stars \citep{BO2009, Bolmont2012, DamianiLanza2015, Ferraz-Mello2015}, it thus becomes mandatory to take into account their potential strong variations over orders of magnitude as a function of stellar age using their best available ab-initio modeling. 
In this work, we thus propose to replace the currently used constant values of stellar tidal dissipation calibrated on observations \citep[e.g.][]{Hansen2010, Hansen2012} by the new values recently computed for convective enveloppes of low-mass stars as a function of stellar mass, age and rotation \citep{Mathis2015a, Mathis2015b} based on the theoretical work by \citet{Ogilvie2013} using realistic stellar grid models \citep{Siess2000}. 
This will allow us for the first time to unravel the effects of the rotation and tidal dissipation history of stars on the evolution of close-in planets. 
In Sec. \ref{model1}, we describe the implementation of the new tidal model in our state-of-the-art celestial mechanics code \citep{Bolmont2011, Bolmont2012}. 
In Sec. \ref{orb_evol}, we explore the orbital evolution of planets of different masses (from Earth mass to Jupiter mass) around various low-mass K-,G-,F-type host stars. 
In Sec. \ref{star_rot_evol}, we examine and discuss the impact of their orbital evolution on the stellar rotational history in the context of asteroseismic constraints which have been recently obtained \citep[e.g.][]{Gizonetal2013,Ceillieretal2016}. 
Finally, in Sec. \ref{Discussion}, we present the conclusions of our study and discuss its perspectives.


\section{Model description}
\label{model1}

\subsection{Tidal model}
\label{tide_equ}

\subsubsection{Matching two different tidal formalisms}
\label{tidal_model_1}

The tidal model we present here is a first step in reconciling two different tidal formalisms.
On the one hand, we base our work on the same tidal model as in \citet{Bolmont2011, Bolmont2012, Bolmont2015}, which computes the orbital tidal evolution of planetary systems and takes into account the evolution of the star. 
The model is an equilibrium tide model \citep[following][]{Alexander1973, Mignard1979, Hut1981, EKH1998}, which is very practical to compute orbital evolution because it works for all eccentricities \citep{Leconte2010} and allows for fast computation.
The dissipation of the kinetic energy of the tides inside the star is usually taken to be constant throughout the system's evolution.\\

On the other hand however, as a star evolves, its radius changes,  mainly on the PMS, a radiative core appears and gets bigger while the convective enveloppe becomes shallower for stars from $0.45~\Msun$ to $1.4~\Msun$. 
All these structural changes have a strong influence on the way the star dissipates tidal energy \citep[e.g.][]{Zahn1966, Zahn1975, Zahn1977, OgilvieLin2007, RMZ2012, Mathis2015a, Mathis2015b} and this should be taken into account in tidal calculations. 

Besides, by definition, the equilibrium tide does not take into account the dynamical tide \citep{Zahn1975, OgilvieLin2007}. 
The equilibrium tide corresponds only to the large-scale flows induced by the hydrostatic adjustment of the star's structure because of the presence of the companion \citep[][]{Zahn1966, RMZ2012}. 
It is mainly dissipated in the convective enveloppe of low-mass stars by the turbulent friction applied by convection \citep[][]{Zahn1977, Zahn1989}. 
However, both in convection and radiation zones, the equilibrium tide is not solution of the hydrodynamics equation \citep[e.g.][]{Zahn1975, OgilvieLin2004, Ogilvie2013} and it must be completed by the so-called dynamical tide. 
In convective regions, the dynamical tide is constituted of inertial waves, which are driven by the Coriolis acceleration, that are excited when the tidal frequency $\omega \in [-2\Os, 2\Os]$, where $\Os$ is the rotation frequency of the star \citep[][]{OgilvieLin2007}. 
If the system is coplanar and the perturber is on a circular orbit, the tidal frequency is simply $\omega\equiv2(n-\Os)$, where $n$ is the orbital frequency. 
This gives the following condition on the orbital period of the planet: $P_{\rm orb} > 1/2 P_{\star}$ for the excitation of tidal inertial waves, where $P_{\star}$ is the rotation period of the star. 
In this context, \citet{Bolmont2012} studied the tidal evolution of planets around evolving stars taking into account the radius evolution of the star and its impact on the tidal evolution of planets. 
The planets were chosen to be initially close to the corotation radius (defined as the orbital distance for which $n = \Os$). 
Consequently, the planets considered in \citet{Bolmont2012} should in fact raise inertial waves in the convective enveloppe of their host star. 
Therefore, a new dissipation, which adds to the one of the equilibrium tide, must be taken into account because of the turbulent friction applied by convective eddies on tidal inertial waves. 
In stellar radiation zones, the dynamical tide is constituted by gravito-inertial waves, in more explicit terms internal gravity waves which are influenced by the Coriolis acceleration \citep[][]{Zahn1975, Terquemetal1998, OgilvieLin2007, Ivanovetal2013, AuclairDesrotouretal2015}. 
In some cases, their dissipation compete with those of the equilibrium tide and inertial waves in convective regions \citep[e.g.][]{GD1998,OgilvieLin2007}, especially when massive companions excite high-amplitude waves that may non-linearly break at the center of K- and G-type stars \citep[][]{BarkerOgilvie2010, Barker2011, Guillotetal2014}. 
Tidal dissipation rates and torques are highly dependent on tidal frequency \citep[e.g.][]{OgilvieLin2004, OgilvieLin2007, AuclairDesrotouretal2015}; therefore, they vary over several orders of magnitude as a function of the excitation frequency. 
This has strong consequences on tidal dynamics \citep[][]{WS1999, AuclairDesrotouretal2014} with possible erratic variations of orbital and rotational properties, which are not treated by "equilibrium tide"-like models. 
Solving hydrodynamics equations for tidal flows and waves must thus be done in the Fourier spectral space. 
This requires to expand tidal potential/force and perturbing function on Fourier series \citep[][]{Kaula1964, Zahn1977, MLP09, EM2013}, with a number of mandatory modes (and frequencies) that strongly increases with orbital eccentricity and inclination, and obliquity \citep[][]{S2008}. 
Then, computation of tidal dynamics can become very heavy compared to "equilibrium tide"-like models.\\   

This is the reason why we adopt several simplifications of the problem in this first work where we wish to take into account the impact of the history of tidal dissipation in stars on the orbital dynamics of close-in planets. 
First, we choose to focus only on tidal dissipation in the external convective enveloppe of low-mass stars hosting planets \cite[][]{Ogilvie2013,Mathis2015a,Mathis2015b}. 
Therefore, we take into account the dissipation of tidal inertial waves when $P_{\rm orb} > 1/2 P_{\star}$ and of the equilibrium tide otherwise. 
Next, we simplify the treatment of the frequency-dependence of the problem adopting two assumptions. 
On the one hand, we focus on binary systems constituted by a planet on coplanar circular orbit around its evolving host star. 
The problem is thus only dependent on the main tidal frequency $\omega\equiv2(n-\Os)$ introduced in the previous paragraph. 
We consider that the only tide driving orbital evolution is the stellar tide (meaning that the planet is synchronized with zero obliquity\footnote{For a close-in planet, the evolution time scale of its rotation period and obliquity is very short. We therefore expect the planets we consider here to be synchronized and with a null obliquity very early in the evolution \citep[e.g.,][]{Leconte2010}.}). 
On the other hand, instead of taking into account the full frequency-dependence of tidal dissipation, we choose to adopt the frequency-averaged approach first introduced by \cite{Ogilvie2013} and applied to low-mass stars by \cite{Mathis2015a, Mathis2015b}. 
The advantage of this method is that it provides us relevant orders of magnitude for the dissipation as a function of structural and dynamical stellar parameters \citep[i.e. their mass, age, corresponding mass and radius of the convective enveloppe, and rotation;][]{Mathis2015a, Mathis2015b} even if it filters out the potential variation of the dissipation over a narrow range of frequencies \citep[e.g.][]{OgilvieLin2004, OgilvieLin2007, AuclairDesrotouretal2015}. The corresponding values of averaged dissipation will be here obtained for a range of stellar masses ($0.4$ to $1.4~\Msun$) with a metallicity of $Z=0.02$ and a mixing length parameter (scaled on the local pressure height scale) $\alpha=1.605$ calibrated on the Sun using grids of realistic models of low-mass stars computed by \citet{Siess2000} with the STAREVOL code.

If the system is coplanar and the orbit of the planet is circular, the frequency averaged tidal dissipation \citep[equation B13 of][]{Ogilvie2013} is given by:
\begin{eqnarray}\label{freq_av_diss}
\lefteqn{\left<{\mathcal D}\right>_{\omega}=\int^{+\infty}_{-\infty} \! {\rm Im} \left[k_2^2(\omega)\right] \,\frac{\mathrm{d}\omega}{\omega} = \frac{100 \pi}{63} \epsilon^2 \left(\frac{\alpha^5}{1-\alpha^5}\right)\left(1-\gamma\right)^2}\\
&&\times\left(1-\alpha\right)^4\left(1+2\alpha+3\alpha^2+\frac{3}{2}\alpha^3\right)^2\left[1+\left(\frac{1-\gamma}{\gamma}\right)\alpha^3\right]\nonumber\\
&&\times\left[1+\frac{3}{2}\gamma+\frac{5}{2\gamma}\left(1+\frac{1}{2}\gamma-\frac{3}{2}\gamma^2\right)\alpha^3-\frac{9}{4}\left(1-\gamma\right)\alpha^5\right]^{-2}\nonumber
\end{eqnarray}
with
\begin{equation}\label{param_freq_av_diss}
\alpha=\frac{R_{\rm c}}{R_{\star}}\hbox{,}\quad\beta=\frac{M_{\rm c}}{M_{\star}}\quad\hbox{and}\quad\gamma=\frac{\alpha^3\left(1-\beta\right)}{\beta\left(1-\alpha^3\right)}<1.
\end{equation}
We introduce the mass ($M_{\star}$) and the radius ($R_{\star}$) of the star, the mass ($M_{\rm c}$) and radius ($R_{\rm c}$) of its radiative core, and its normalized rotation $\epsilon \equiv \left( \Os/\sqrt{\G\Ms/\Rs^3}\right) = (\Os/\Omega_{\star, {\rm c}})$, $\Omega_{\star, {\rm c}}$ being the critical angular velocity of the star. 
In the case of a coplanar orbit, the expansion of the tidal force on spherical harmonics reduces to the quadrupolar mode $\left(l=2,m=2\right)$. 
We introduce the corresponding Love number $k_2^2$ giving the ratio between the perturbation of the gravitational potential induced by the presence of the planetary companion and the tidal potential evaluated at the stellar surface. 
Its imaginary part $\mathrm{Im}\left [k_2^2(\omega)\right ]$, which provides us a direct quantification of tidal dissipation, can be expressed in terms of the tidal quality factor $Q_2^2(\omega)$ or the tidal angle $\delta_2^2(\omega)$ \citep[e.g.][]{RMZ2012,EM2013}:
\begin{equation}\label{Q_Imk_delta}
Q_2^2(\omega)^{-1} = \mathrm{sgn}(\omega)\left | k_2^2(\omega)\right |^{-1} \mathrm{Im}\left [k_2^2(\omega)\right ] = \sin\left [2\delta_2^2(\omega)\right ].
\end{equation}
This allows us to define an equivalent quality factor $\overline{Q}$ and an equivalent modified quality factor $\overline{Q'}$ as defined by \citet{OgilvieLin2007} and \citet{Mathis2015a} and the corresponding equivalent tidal angle $\overline{\delta}$\footnote{We point out here that equivalent quality factors ${\overline{Q'}}$ and ${\overline Q}$, which are proportional to the inverse of the frequency-averaged dissipation $\left<{\rm Im} \left[k_2^2(\omega)\right]\right>_{\omega}$, where $\left<...\right>_{\omega}=\int_{-\infty}^{\infty}...{\mathrm{d}\omega}/{\omega}$, are not equivalent to potentially defined frequency-averaged quality factors $\left<Q'\left(\omega\right)\right>_{\omega}$ and $\left<Q\left(\omega\right)\right>_{\omega}$. In this framework, the relevant physical quantity being $\left<{\rm Im} \left[k_2^2(\omega)\right]\right>_{\omega}$, we prefer to define directly equivalent quality factors from it.}:
\begin{equation}\label{Q_Q'}
\frac{3}{2\overline{Q'}} = \frac{k_2}{\overline{Q}} = \sin\left [2\overline{\delta}\right ] = \int_{-\infty}^{\infty } \mathrm{Im}[k_2^2(\omega)] \frac{\dd \omega}{\omega} = \left<{\mathcal D}\right>_{\omega},
\end{equation}
where $k_2$ is here the usual quadrupolar Love number, which evaluates the non-dissipative hydrostatic elongation of the star in the direction of the planet. 

Following \citet{Mathis2015a}, we can decouple the rotation part from the purely structural part in Eq. (\ref{freq_av_diss}). 
Therefore, for a star rotating at a fixed angular velocity $\Os$, we can define a complementary frequency-averaged dissipation at fixed rotation:
\begin{equation}\label{freq_av_diss_struc}
\langle\mathcal{D}\rangle_\omega^{\Os} = \epsilon^{-2} \langle\mathcal{D}\rangle_\omega = \epsilon^{-2} \langle \mathrm{Im}\left[k_2^2(\omega)\right] \rangle_\omega,
\end{equation}
which allows us to isolate the dependence of the dissipation on the internal structure of the star. 
It can also be expressed in terms of $\hat{\epsilon} = \left( \Os/\sqrt{\G\Msun/\Rsun^3}\right) = \Os/\Omega_{\odot, {\rm c}}$, where $\Omega_{\odot, {\rm c}}$ is the critical angular velocity of the Sun. 
We choose here to use $\hat{\epsilon}$ and Eq. (\ref{freq_av_diss_struc}) becomes:
\begin{equation}\label{freq_av_diss2}
\langle\mathcal{\hat{D}}\rangle_\omega^{\Os} = \hat{\epsilon}^{-2} \langle\mathcal{D}\rangle_\omega = \left(\frac{\Ms}{\Msun}\right)^{-1}\left(\frac{\Rs}{\Rsun}\right)^{3} \langle\mathcal{D}\rangle_\omega.
\end{equation}
As a consequence, we can introduce an equivalent structural tidal quality factor $\Qs$ defined as a function of the equivalent tidal quality factor $\overline{Q'}$
\begin{equation}\label{freq_av_diss3}
\frac{3}{2\overline{Q'}} = \hat{\epsilon}^{2} \frac{3}{2\Qs},
\end{equation}
so that when the spin of the star increases, $1/\Q$ increases, thus $\Q$ decreases and the dissipation increases.\\

The orbital dynamics code we use has been written in the framework of the constant time lag model.
This model is very practical to compute orbital evolution, we thus want to continue using it. 
However, we now allow the dissipation $\oss$ \citep[normalized bulk dissipation per unit mass, as defined in][]{Hansen2010} to vary while the star evolves.
The dimensionless $\oss$ is linked with the dissipation factor $\sss$ and the time lag by the following formula:
\begin{equation}\label{ss}
\oss = \sss/\sigma_0 = k_2^2 \Delta \tau_\star \frac{2\G}{3\Rs^5}\frac{1}{\sigma_0},
\end{equation}
where $\sigma_0 = \sqrt{\G/(\Msun\Rsun^7)}$ \citep{Hansen2010, Bolmont2015}. 
In \citet{Bolmont2012}, $\oss$ was considered constant throughout the evolution, which meant that $k_2^2 \Delta \tau_\star$ was actually evolving along with the radius of the star.

In order to incorporate the formalism from \citet{Mathis2015b} in the framework of the constant time lag model, one needs to express the time lag $\Delta \tau$ as a function of the equivalent lag angle $\overline{\delta}$. 
As discussed by \cite{Leconte2010}, this is not straightforward in the general case. 
Indeed, introduced equivalent tidal angle and quality factor are related to frequency-averaged quantities computed in the Fourier space while the constant time lag model is a model using quantities in real space such as bodies' position and velocity. 
We first do the approximation that the lag angle $\overline{\delta}$ is small so that $\sin\left[2\overline{\delta}\right] \approx 2\overline{\delta}$. Besides, 
\begin{equation}
2\overline{\delta} = \omega \Delta \tau,
\label{deltaTauDef}
\end{equation}
where $\omega$ is the tidal frequency. In the general case of an eccentric orbit, several tidal frequencies would have to be taken into account and it would become necessary to have several terms associated with each Fourier mode. 
However, in the simplest case considered here of a circular orbit, there is a unique frequency $\omega\equiv2|n-\Os|$. Consequently, the time lag is given by:
\begin{equation}\label{delta_tau}
\Delta \tau_\star = \frac{\overline{\delta}}{|n-\Os|} = \frac{3}{4\Q |n-\Os|} = \frac{3\hat{\epsilon}^{2}}{4\Qs |n-\Os|}.
\end{equation}
We recover the fact that when $\Qs$ is small, $\Delta \tau_\star$ and $\oss$ are big and consequently the tidal evolution timescales are shorter.
As in \citet{Mathis2015b}, we see that the higher the spin of the star, the higher the dissipation $1/\Q$ or $\oss$. 

Note that for $n=\Os$, there is a non-physical singularity intrinsic to the definition of the constant time lag model given in Eq. (\ref{deltaTauDef}) which is here the price of a lighter computation. 
It would be possible to avoid it as soon as the full complex frequency-dependence of tidal dissipation will be taken into account. 
Then, it would be possible to define a spectral time lag $\Delta \tau\left(\omega\right)$ computed directly from $\mathrm{Im}\left [k_2^2(\omega)\right]$. 
However, the simple and compact equations derived by \cite{Hut1981} would have to be abandoned in favor of a fully spectral treatment of dynamical equations \citep[][]{MLP09, RMZ2012, Ogilvie2014} needing potential heavy computations. 
In this first qualitative work, to be able to continue to use the compact formalism, we choose to smooth out the artificial singularity by using the following formula:
\begin{equation}\label{delta_tau_code}
\Delta \tau_\star = \frac{\overline{\delta}}{\max{\left[|n-\Os|, \rho\right]}},
\end{equation}
where $\rho$ is a regularization value to avoid the singularity. 
$\rho$ has the dimension of a frequency.
We tested $\rho = 10^{-8}$~s$^{-1}$ and $\rho = 10^{-5}$~s$^{-1}$ and qualitative differences appear for the most extreme cases initially close to the corotation radius: a planet initially at corotation can fall on the star in one case and survive in the other. 
However for most cases, the evolution is qualitatively similar, although not exactly quantitatively similar.

\subsubsection{Tidal secular evolution}

Thanks to this method linking the equivalent modified quality factor to the time lag (or to the dissipation factor $\sss$), we are able to simulate the tidal evolution of planets using the constant time lag formalism \citep{Mignard1979, EKH1998, Bolmont2011}.
Taking into account here only the stellar tide, the secular tidal evolution of the semi-major axis $a$ is given by: 
\begin{equation}\label{Hansena}
\frac{1}{a}\frac{\dd a}{\dd t} = - \frac{1}{\Ts}\Big[Na1(e)-\frac{\Os}{n}Na2(e)\Big],
\end{equation}
where the dissipation timescale $\Ts$ is defined as
\begin{equation}
\label{Tp}
\Ts = \frac{1}{9}\frac{\Ms}{\Mp(\Mp+\Ms)}\frac{a^8}{\Rs^{10}}\frac{1}{\sigma_{\star}}
\end{equation}
and depends on the stellar mass $\Ms$, its dissipation $\sigma_{\star}$ and the planet mass $\Mp$. $Na1(e)$ and $Na2(e)$ are eccentricity-dependent factors, which are valid even for very high eccentricity \citep{Hut1981}:
\begin{align*}
Na1(e) &= \frac{1+31/2e^2+255/8e^4+185/16e^6+85/64e^8}{(1-e^{2})^{15/2}},\\
Na2(e) &= \frac{1+15/2e^2+45/8e^4+5/16e^6}{(1-e^{2})^{6}}.
\end{align*}
In the present case, as the orbits we consider are circular, $Na1(e) = Na2(e) = 1$.

In our model, the planets are driven by the dynamical tide when $P_{\rm orb} > 1/2 P_\star$ and by the equilibrium tide when $P_{\rm orb} < 1/2 P_\star$.
When the equilibrium tide is driving the evolution, the dissipation factor is taken to be the normalized bulk dissipation per unit mass $\oss$ for a 1~Gyr star and a tidal period of $1$~day given in \citet{Hansen2012}.
This equilibrium tide factor is given in Table \ref{tab:param:star} for the stars considered in this work.
When the dynamical tide is driving the evolution, the dissipation factor is obtained using Equations \ref{ss} to \ref{delta_tau_code}.



%

\subsubsection{Stellar wind}\label{wind_stuff}

We also take into account the spin down of the star due to the stellar wind. 
As in \citet{Bouvier1997}, we assume that the stars considered here rotate as solid bodies, although more recent work has included the effect of internal differential rotation between the radiative core and the convective enveloppe \citep[e.g.,][]{Bouvier2008,GalletBouvier2013,Penevetal2014,GalletBouvier2015}. 
The wind braking processes at work for a Sun-like star are the same as for other low-mass stars \citep{Matt2015} so we consider the same wind parametrization for all stars.
We use here the torque formula from \citet{Bouvier1997} with updated estimates for the saturation spin $\omega_{sat}$ from \citet{Matt2015}.
In this work, $M_\star$ is held constant, and the effect of mass loss (through processes like stellar winds) on the internal structure of the star is considered negligible.

Our calculations begin during the stellar Pre Main Sequence (PMS). 
As in \citet{Bolmont2012} we use the ``disk locking'' parametrization \citep[e.g.,][]{Bouvier1997, Rebull2004, Rebull2006, Edwards1993, ChoiHerbst1996}, which consists in assuming that the rotation period of the star remains at an initial constant value for a given time (hypothesized to be associated with the time of disk dissipation).
After this time, the spin of the star evolves from this initial value depending on its radius contraction rate and the influence of the stellar wind.
We thus start our calculations at the moment of ``disk dispersal'' and we consider that the planets are fully formed at that point.\\

One goal of the present work is to determine how different stellar spin and dissipation histories influence the star-planet tidal interaction. 
To this end, we consider three different stellar masses: $0.6$, $1.0$ and $1.2~\Msun$ and consider an initial spin rate of $P_{\star, 0} =1.2$~day, which corresponds to the fast enveloppe of the observed stellar spin distribution \citep{Bouvier1997}.
In Section \ref{init_spin}, we consider other initial spin rates: $P_{\star, 0} = 8$~day, which corresponds to the slow enveloppe of the observed stellar spin distribution and an intermediate initial spin rate: $P_{\star, 0} = 3~$day.

        \begin{figure*}[htbp!]
	\centering
        \includegraphics[width=15cm]{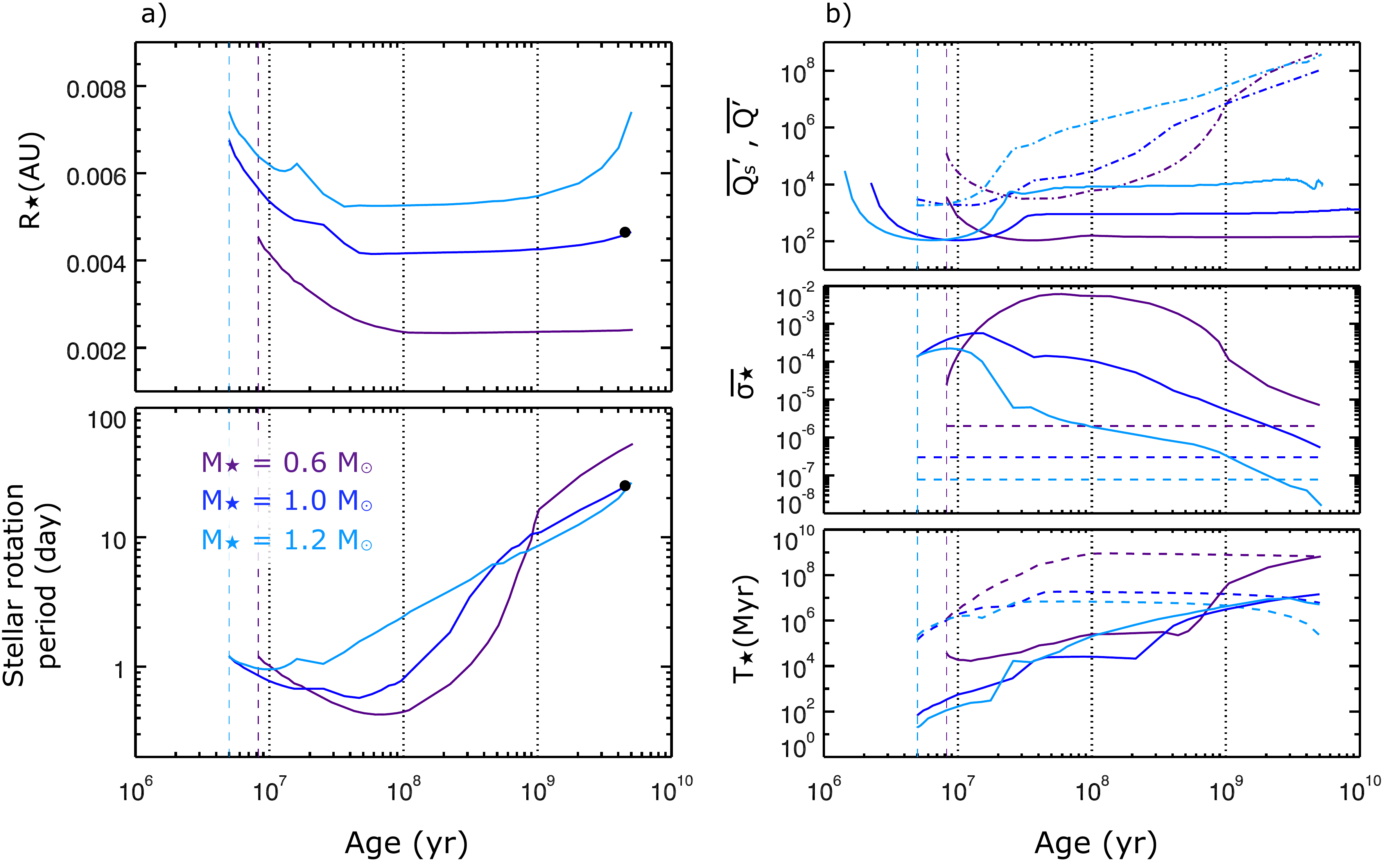}
        \caption{Evolution of radius, rotation period and dissipation of the different stars for a $1~\Mearth$ planet at $1000$~AU, and timescales evolution. a) Top: evolution of the radius of the star. The present radius of the Sun is represented by a black dot. Bottom: evolution of the rotation period of the star. The present rotation period of the Sun is represented by a black dot. b) Top: the equivalent modified quality factors $\Qs$ (full lines) and $\Q$ (dashed-dotted lines). Middle: the corresponding dissipation factor $\oss$ (dynamical tide dissipation in full lines, constant equilibrium dissipation in dashed lines). Bottom: Evolution timescale $T_\star$ for the three stars but for a $1~\Mearth$ planet at $0.026~$AU (dynamical tide evolution in full lines, constant equilibrium tide evolution in dashed lines). The vertical dashed lines correspond to the initial time considered in our simulations. These simulations were done with a regularization value $\rho$ of $10^{-5}$~s$^{-1}$. The initial rotation period of the stars is $1.2$~day.}
        \label{spin_Q_sigma_for_Ms}
        \end{figure*}

As in \citet{Bolmont2012}, we consider both the influence of tides and the stellar wind on the rotation of the star.
The expression for the angular momentum loss rate is \citep{Kawaler1988, MacGregorBrenner1991, Bouvier1997}: 
\begin{align} \label{truc}
\begin{split}
  \frac{1}{J}\frac{\dd J}{\dd t} &= \frac{-1}{J}K\Omega_\star^\mu \omega_{sat}^{3-\mu}\left(\frac{R_{\star}}{\Rsun}\right)^{1/2}\left(\frac{M_{\star}}{\Msun}\right)^{-1/2} \\
 & + \frac{1}{J}\frac{h}{2T_\star}\left[ No1(e)-\frac{\Omega_{\star}}{n}No2(e)\right], 
\end{split}
\end{align}
where $h$ is the orbital angular momentum, $n$ is the mean orbital angular frequency, $T_\star$ is the stellar dissipation timescale defined in Eq. (\ref{Tp}), and the functions $No1$ and $No2$ are defined as:
\begin{align*}
No1(e) &= \frac{1+15/2e^2+45/8e^4+5/16e^6}{(1-e^{2})^{13/2}},\\
No2(e) &= \frac{1+3e^2+3/8e^4}{(1-e^{2})^{5}}.
\end{align*}
In the present case, as the orbits we consider are circular, $No1(e) = No2(e) = 1$.
 
Here $K$, and $\omega_{sat}$ are parameters of the model from \citet{Bouvier1997}. 
We use the value of $K=1.7 \times 10^{47}$~cgs for all stars and we choose $\omega_{sat}$ from \citet{Matt2015} so that: 
\begin{equation*}
\begin{cases}
 \omega_{sat} = 3.1~\Omega_\odot, & \text{for } \Ms = 0.6~\Msun  \\
 \omega_{sat} = 9~\Omega_\odot, & \text{for } \Ms = 1.0~\Msun \\
 \omega_{sat} = 31~\Omega_\odot, & \text{for } \Ms = 1.2~\Msun
  \end{cases}\,.
\end{equation*}
\citet{Bouvier1997} showed that for fast rotators ($\Omega_\star>\omega_{sat}$), $\mu = 1$ and for slow rotators ($\Omega_\star<\omega_{sat}$), $\mu = 3$.

\subsubsection{Evolution of stellar rotation period and dissipation}

Figure \ref{spin_Q_sigma_for_Ms} shows the evolution of the radius and rotation period of the star, as well as the structural equivalent quality factor $\Qs$ and its counterpart $\Q$ (cf Eq. \ref{freq_av_diss3}), and the normalized dissipation factor $\oss$ for the three different stars considered here: $\Ms = 0.6~\Msun$, $1~\Msun$ and $1.2~\Msun$.
All stars have an initial rotation period of $1.2$~day.
For the last panel of Fig. \ref{spin_Q_sigma_for_Ms}b), we simulated the evolution for two cases.
Either the evolution is driven following our model (full lines) or the evolution is solely driven by the equilibrium tide with a constant dissipation \citep[dashed lines, as in][]{Bolmont2012}.
Table \ref{tab:param:star} shows the stellar parameters used in this work.

\begin{table}[htbp]
\begin{center}
\caption{Stellar parameters.}
\vspace{0.1cm}
\begin{tabular}{c|c|c|c}
Mass & Initial time & Equilibrium tide  & Initial rotation \\
($\Msun$) & (Myr) & dissipation $\oss$ & (day) \\
\hline
0.6 & 8.2 & $2.1\times 10^{-6}$ & 1.2/ 3/ 8 \\
1 & 5 & $3\times 10^{-7}$ & 1.2/ 3/ 8 \\
1.2 & 5 & $7.8\times 10^{-8}$ & 1.2/ 3/ 8 \\
\end{tabular} 
\label{tab:param:star} 
\end{center}
\end{table}

These evolutions were calculated for a $1~\Mearth$ planet orbiting the different stars at a semi-major axis of $1000$~AU, which means that the influence of the planet on the evolution of the star is negligible. 
Besides, the planet is so far away from the star that it does not actually tidally evolve in the 5~Gyr timescale of the simulation.
Consequently, this allows us to see the variation of the dissipation in the star due to its structural evolution.
Besides, in this configuration where the planet is very far away, the planet following our model is always evolving due to the dynamical tide ($P_{\rm orb}$ is always higher than $1/2 P_\star$). 
The tidal frequency can be here approximated by $\sim 2\Os$.


Table \ref{tab:param:star} also shows the initial time we consider for the beginning of our simulations.
This initial time corresponds to the time of disk dissipation, when the disk no longer has an influence on the dynamical evolution of the planets. 
For $\Ms = 1~\Msun$ and $\Ms = 1.2~\Msun$, the initial time is taken to be 5~Myr, however for $\Ms = 0.6~\Msun$ it is taken to be $8.2$~Myr.
The lower the mass of the star, the later the radiative core appears.
For a star of mass $\Ms = 0.6~\Msun$, this happens at an age of $8.2$~Myr.
Before the apparition of the radiative core, the inertial waves propagate in a fully convective sphere and cannot reflect in a way that leads to the formation of sheared waves attractors that may lead to strong dissipation \citep{OgilvieLin2004,OgilvieLin2007}; the dissipation is therefore very weak and the equivalent modified quality factor very high.


Figure \ref{spin_Q_sigma_for_Ms} shows that the spin of the stars evolves accordingly with the observations \citep{Matt2015}.
Indeed, for $\Ms = 1~\Msun$, we reproduce today's Sun rotation period (the dot in Fig. \ref{spin_Q_sigma_for_Ms}). 
For $\Ms = 1.2~\Msun$, the rotation at $5$~Gyr is comparable to that of the Sun, while for $\Ms = 0.6~\Msun$ it is slower.

Figure \ref{spin_Q_sigma_for_Ms} also shows that the values of the dissipation $\oss$ in our model are in agreement with the equivalent modified quality factor $\Qs$ from \citet{Mathis2015b}.
Indeed, during the PMS stage the dissipation of the high mass stars is higher than that of the low mass stars ($\Qs$ decreases with mass) but during the MS stage the dissipation of the low mass stars is higher than the dissipation of the high mass stars ($\Qs$ increases with mass).
The dissipation factor $\oss$ evolves on the one hand due to the changes of the structural equivalent modified quality factor $\Qs$, and on the other hand due to the changes in the spin of the star (through the parameter $\hat{\epsilon}$).
Figure \ref{spin_Q_sigma_for_Ms} also shows the evolution of $\Q$, which encompasses both effects.

\citet{Ferraz-Mello2015} gives the value of $Q$ for a number of observed planetary systems. 
Among them, the CoRoT candidate LRc06E21637 ($\Ms = 0.93~\Msun$) allows us to compare our value of the tidal quality factor $\overline{Q}$ with the value of $Q = 6\times10^6$ given by \citet{Ferraz-Mello2015}.
The orbital period of the brown-dwarf companion is $P_{\rm orb} = 5.81$~day and the rotation period of the star is $P_\star = 8.95$~day, so the evolution of this system is driven by the dynamical tide.
If the star is on the main sequence, our model gives a value of the structural tidal quality factor of $\Qs = 10^{3}$. 
From the rotation period of the star and Eq. \ref{freq_av_diss3}, we can thus infer a value of $\Q$ of $5.9\times10^{6}$.
Assuming that the quadrupolar Love number of this star is $k_2 = 0.03$ and using Eq. \ref{Q_Q'}, this gives us a value of $\overline{Q} = 1.18\times10^5$, which is 50 times lower than the value of \citet{Ferraz-Mello2015}.
Given that \citet{Ferraz-Mello2015} does not take into account the dynamical tide, this should be the reason why we find a higher dissipation (lower $\overline{Q}$).

Let us consider $\Ms = 1~\Msun$.  
At first $\Qs$ decreases and the rotation increases which has the effect of increasing the dissipation $\oss$. 
Then, from a age of 10~Myrs, $\Qs$ increases but the rotation still increases sufficiently so that the dissipation increases for about another $\sim5$~Myr.
From $\sim15$~Myr to $\sim35$~Myr (i.e., when the star reaches the MS), the dissipation $\oss$ decreases because the spinning up of the star brakes down and $\Qs$ still increases. 
Then, from $\sim35$~Myr to the end of the simulation, $\Qs$ is constant and the dissipation $\oss$ decreases slowly due to the spinning down of the star. 

For a star of $\Ms = 0.6~\Msun$, the increase of the dissipation is more pronounced due to the fact that the star spins up longer and faster and also that $\Qs$ does not evolve as much as for the Sun-like star.
However, for a star of $\Ms = 1.2~\Msun$, the increase of dissipation is less pronounced because it does not spin-up as much as the Sun-like star and $\Qs$ decreases by almost 2 orders of magnitude when reaching its minimum in the PMS.  

Finally, Figure \ref{spin_Q_sigma_for_Ms}b also shows the evolution of the tidal timescale $\Ts$ (Eq. \ref{Tp}) for a $1~\Mearth$ planet at $0.026$~AU (fixed semi-major axis) around the different stars.
The timescale $\Ts$ was computed for two cases: due to the dynamical tide (regardless of the domain of validity, represented in full lines in Fig. \ref{spin_Q_sigma_for_Ms}b) and due to the equilibrium tide (dashed lines). 
For this last panel, we do not consider the case of the planet at $1000$~AU because its tidal evolution timescale would be much larger than the age of the universe. 
This last panel of Figure \ref{spin_Q_sigma_for_Ms}b is to show the typical evolution of the tidal timescale with the structural evolution of the star. 
For all stars, the equilibrium tide evolution timescale (dashed lines) increases with time due to the shrinking of the radius. 
It decreases with increasing stellar mass, due to the higher stellar radius for higher masses.
The dynamical tide evolution timescale also increases with time, but this time its evolution is not solely linked to the evolution of the radius, but also the evolution of $\oss$ (or $\Q$, or $\Qs$ and $\Os$).
The evolution timescales reaches with a plateau more or less pronounced corresponding to the moment where the structural equivalent modified quality factor $\Qs$ becomes constant.

The comparative evolution of the timescales depending on the mass of the star is not straightforward because it depends on a lot of parameters: the structural evolution of the star, through the radius and $\Qs$ and also on the spin.
Nonetheless, we can see that in the beginning and in the end of the stellar evolution considered here, $\Ts$ decreases when increasing stellar mass.  
In the beginning of the evolution, $\Ts$ is smaller for $1.2~\Msun$ because of both the higher dissipation $\oss$ (or lower $\Q$) and the higher radius, while from 4~Gyr to 5~Gyr, despite a smaller dissipation $\oss$, $\Ts$ is smaller for $1.2~\Msun$ only because the radius is much bigger than the $0.6$ and $1~\Msun$ stars.


\section{Orbital tidal evolution}\label{orb_evol}

We therefore investigate how this new model impacts the orbital tidal evolution of planets compared with the standard equilibrium tide model of \citet{Bolmont2012}.
We consider planets from $1~\Mearth$ to $1~\Mjup$ orbiting the three different stars considered here.

\subsection{Evolution of planets orbiting $1~\Msun$}\label{evol_1Msun}

\begin{figure}[htbp!]
        \begin{center}
        \includegraphics[width=9cm]{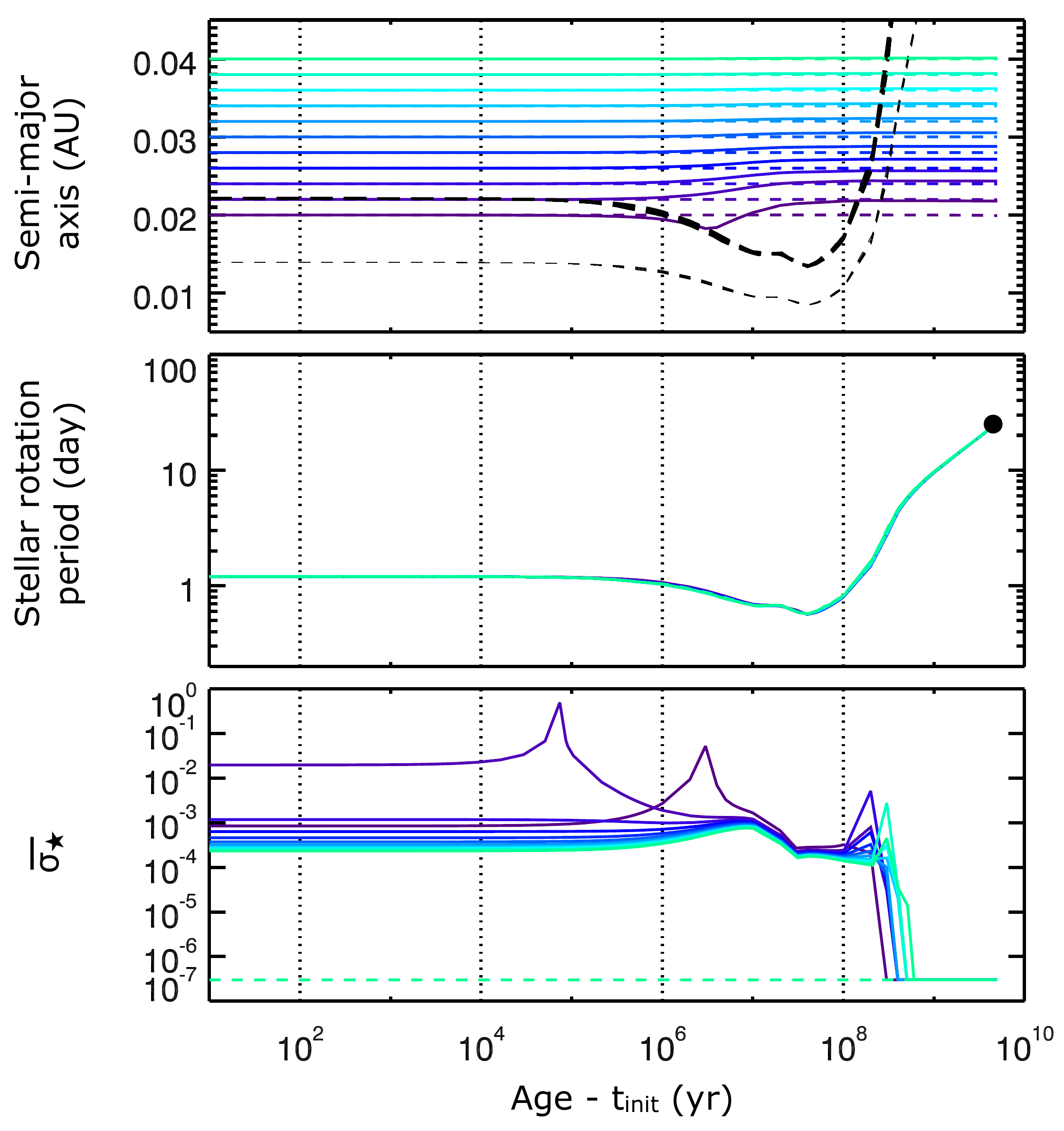}
        \caption{Tidal evolution of a $1~\Mearth$ planet around an evolving $1~\Msun$ star. The evolution calculated with the equilibrium tide model using \citet{Hansen2012}'s dissipation factor (see Table \ref{tab:param:star}) is represented in dashed lines. The evolution calculated with our model of the dynamical tide is represented in full lines. Top: evolution of the semi-major axis of the planets (colors), corotation radius of the star (bold black dashed lines) and the line defining $P_{\rm orb} = 1/2 P_\star$ (thin black dashed lines). Middle: rotation period of the star, the rotation of the Sun at present time is represented with a black dot. Bottom: Evolution of the normalized dissipation factor $\oss$. These simulations were done with a regularization value $\rho$ of $10^{-8}$~s$^{-1}$.}
        \label{comp_dissip_sma_spins_deltat_Mp_1_Ms_1_smooth_1d8}
        \end{center}
        \end{figure}

Figure \ref{comp_dissip_sma_spins_deltat_Mp_1_Ms_1_smooth_1d8} shows the tidal evolution if the system was evolving solely due to the equilibrium tide (dashed lines) and if it was evolving following our model (full lines). 
In the latter scenario, the planet evolves due to the dynamical tide during the first few 100~Myr of its evolution (when $P_{\rm orb} > 1/2 P_\star$).
When the star has spun down sufficiently, so that the orbital period becomes less than half the stellar rotation period, the equilibrium tide takes over and the dissipation $\oss$ decreases to the value of \citet{Hansen2012}.

        \begin{figure*}[htbp!]
	\centering
        \includegraphics[width=15cm]{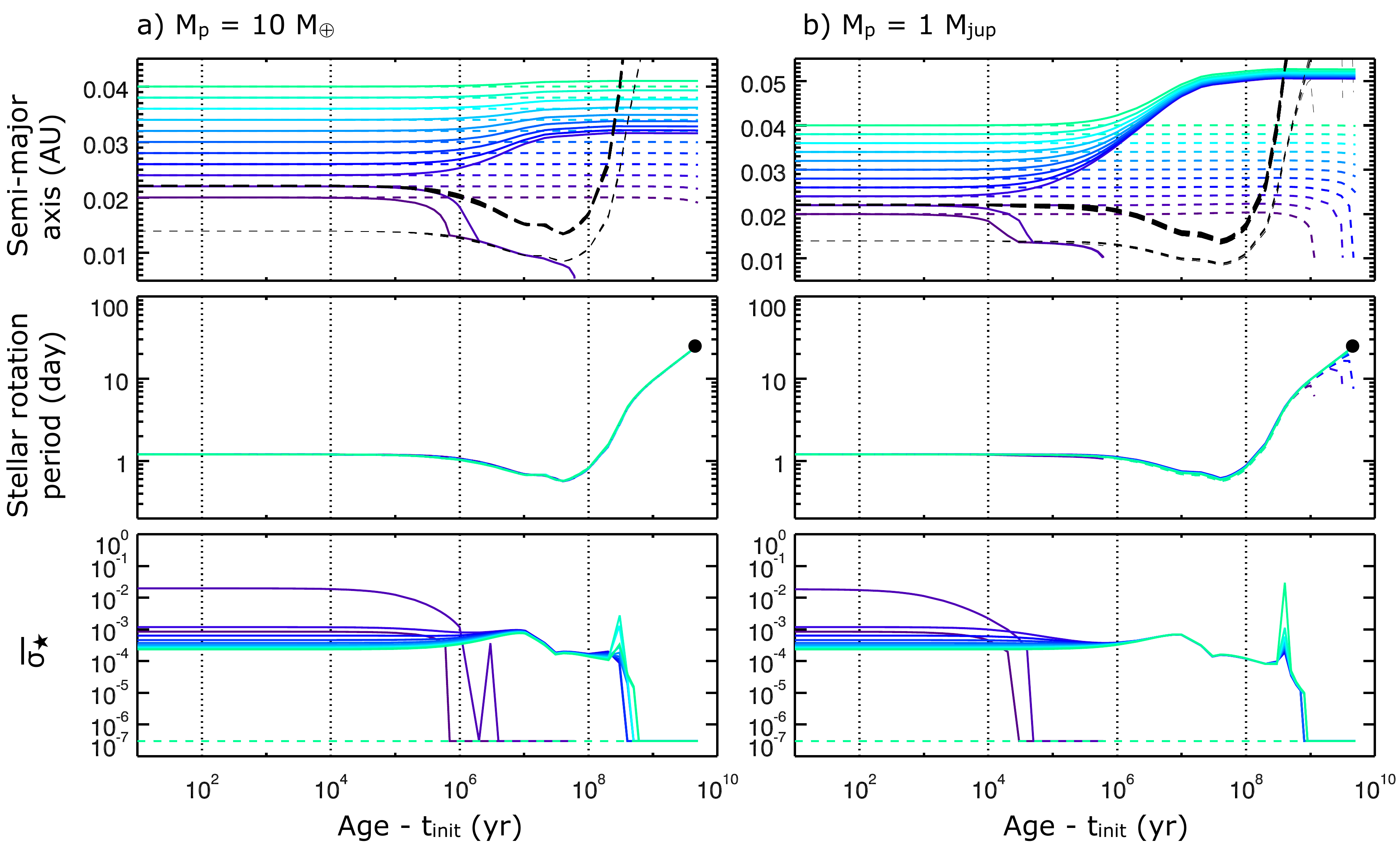}
        \caption{As Figure \ref{comp_dissip_sma_spins_deltat_Mp_1_Ms_1_smooth_1d8} but for a) a planet of $10~\Mearth$ and b) a planet of $\Mjup$.}
        \label{comp_dissip_sma_spins_sigmasbar_Mp_10_Ms_1_smooth_1d8}
        \end{figure*}

For planets closer than $0.03$~AU, the difference between an evolution driven by the equilibrium tide and an evolution driven by the dynamical tide is clearly visible. 
For a planet of $1\Mearth$, due to the very low dissipation, the equilibrium tide do not cause any significant migration.
However, as the dynamical tide is responsible for a dissipation factor higher by several orders of magnitude than the equilibrium tide dissipation, the planet experiences a faster migration.
This means that the planet is much more sensitive to the rotation period evolution of the star: planets interior to the corotation radius will migrate inward, while those exterior to the corotation radius migrate further outwards than before. 

Figure \ref{comp_dissip_sma_spins_deltat_Mp_1_Ms_1_smooth_1d8} shows that the rotation period of the star is not particularly influenced by the planet. 
Except from a small deviation around $t = 1$~Myr due to the passage of the planet through corotation, the spin of the star follows the same evolution for all planets and the rotation tends to that of the Sun at present day \citep[as observed for example for KOIs, see e.g.][]{Ceillieretal2016}.

For the innermost planet, we notice the same kind of behavior that was first observed by \citet{Bolmont2011} of inward migration followed by outward migration as the planet crosses the shrinking corotation radius.
This planet was ``saved'' from falling onto the star by the spinning-up of the star. 
Later on in the evolution, the planet crosses once more the expanding corotation radius. 
However, as the dissipation is lower at that time because of the evolution of $\Qs$ and of the rotation, no significant orbital changes are visible.
As discussed in section \ref{tidal_model_1}, note that when the planet crosses corotation the dissipation increases.
 
In this framework, we investigated the influence of the parameter $\rho$ that smoothes the divergence when $n = \Os$ (Eq. \ref{delta_tau_code}) and for most cases it is negligible. 
In some cases, massive planets located very close to the corotation can either survive and migrate outward when $\rho = 10^{-5}$~s$^{-1}$, or migrate inward and fall onto the star when $\rho = 10^{-8}$~s$^{-1}$.
Apart from these very specific cases, the difference in $\rho$ only influences slightly the final semi-major axis (less than a percent). 


        \begin{figure*}[htbp!]
	\centering
        \includegraphics[width=15cm]{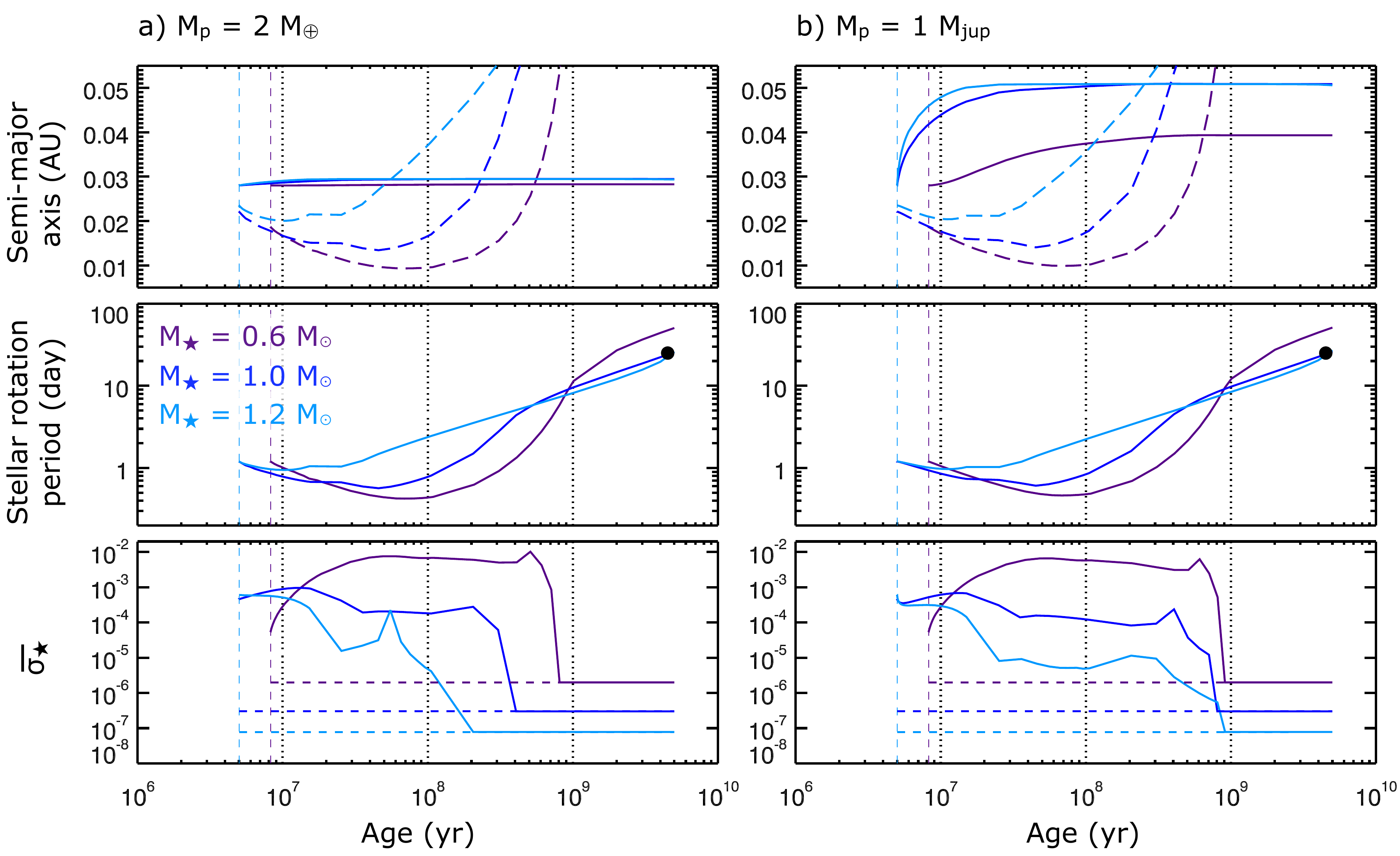}
        \caption{Tidal evolution of a) a $2~\Mearth$ planet and b) a $\Mjup$ planet initially at $0.028$~AU around the different stars. The initial time for the simulations are: 8~Myr for $0.6~\Msun$ and 5~Myr for $1~\Msun$ and $1.2~\Msun$. Top: evolution of the semi-major axis of the planets (full lines), corresponding corotation radius (bold dashed lines). Middle: rotation period of the star, the rotation of the Sun at present time is represented with a black dot. Bottom: Evolution of the normalized dissipation factor $\oss$ (full lines) and the equilibrium tide dissipation factor (dashed lines). These simulations were done with a regularization value $\rho$ of $10^{-8}$~s$^{-1}$.}
        \label{comp_dissip_sma_spins_sigmasbar_Mp_2_318_Ms_smooth_1d8}
        \end{figure*}


When increasing the mass of the planet, the tide raised in the star is bigger and the planets migrate outward farther away.
Figure \ref{comp_dissip_sma_spins_sigmasbar_Mp_10_Ms_1_smooth_1d8}a shows the evolution of a $10~\Mearth$ planet. 
The planets experiencing outward migration migrate from $\sim 0.02$~AU to semi-major axes bigger than $0.03$~AU.
The two planets beginning below the corotation radius migrate inwards. 
The inward migration happens in three steps. 
The first one is due to the dynamical tide which very quickly makes the planet migrate to the limit $P_{\rm orb} = 1/2 P_\star$ (thin black dashed lines in Fig. \ref{comp_dissip_sma_spins_sigmasbar_Mp_10_Ms_1_smooth_1d8}a).
Once the planet reaches this limit, after a few thousand hundred years, the equilibrium tide drives the evolution with a much lower dissipation. 
The planets's inward migration then occurs on a much longer timescale.
However the star continues spinning up so that $P_\star$ decreases on shorter timescales than $P_{\rm orb}$ does. 
When $1/2 P_\star$ becomes smaller than $P_{\rm orb}$, the dynamical tide drives the inward migration once more until $P_{\rm orb} = 1/2 P_\star$.
Consequently, the dissipation in the star jumps back and forth from the dynamical value to the equilibrium value and the planet stays on the limit $P_{\rm orb} = 1/2 P_\star$ for a few million years.
Figure \ref{comp_dissip_sma_spins_sigmasbar_Mp_10_Ms_1_smooth_1d8}a shows this behavior but the sampling of our outputs do not allow to see the dissipation jumping more than once.
In the end, the planet migrates inward sufficiently due to the equilibrium tide to escape this blockage and collide with the star in less than 100~Myr.

When increasing even more the mass of the planet, for example to $\Mp = 1~\Mjup$, the planets initially outside the corotation radius migrate even farther away. 
Figure \ref{comp_dissip_sma_spins_sigmasbar_Mp_10_Ms_1_smooth_1d8}b shows that typically for a star with an initial rotation period of $1.2$~day, the planets migrate from $\sim0.02$~AU to $0.05$~AU in a Gyr or so. 
For Jupiter mass planets, there is a huge difference in the evolution following our model including the dynamical tide and the evolution due to the equilibrium tide. 
Planets beginning just outside the corotation radius fall onto the star in a few Gyr when evolving due to the equilibrium tide. 
As the equilibrium tide evolution timescale is very big, the planets are not sensitive to the initial spin of the star.
Their tidal evolution is noticeable only when the star has already spun down significantly.
However, when evolving due to the dynamical tide, they survive and migrate to $0.05$~AU. 

As discussed in \citet{Bolmont2012}, when the planet falls at late ages onto the braking star, it entails an important spinning-up of the rotation. 
However, with our new model, the planets falling onto the star do so very early in the stellar history so that the spin history of the star is not altered significantly (see Sec. \ref{star_rot_evol}).


\subsection{Evolution of planets orbiting other stars}

Figure \ref{comp_dissip_sma_spins_sigmasbar_Mp_2_318_Ms_smooth_1d8} shows the evolution of a planet of $2~\Mearth$ (\ref{comp_dissip_sma_spins_sigmasbar_Mp_2_318_Ms_smooth_1d8}a) and $1~\Mjup$ (\ref{comp_dissip_sma_spins_sigmasbar_Mp_2_318_Ms_smooth_1d8}b) initially orbiting each of the three different stars outside of corotation. 

For low mass planets, the evolution does not depend a lot on the star, they do not migrate very far away from the star. 
For a planet of mass $2~\Mearth$, a difference can be seen between the lowest mass star and the two others. 
Figure \ref{comp_dissip_sma_spins_sigmasbar_Mp_2_318_Ms_smooth_1d8}a shows that the dissipation of the $1~\Msun$ star is very similar to the dissipation of the $1.2~\Msun$ star for the first million year of the evolution (which can also be seen in Fig. \ref{spin_Q_sigma_for_Ms}). 
Consequently, the planets orbiting these two stars undergo a very similar evolution. 
The difference in the dissipation of the $1~\Msun$ star and the $1.2~\Msun$ occurring after one million year is not sufficient to cause a difference in the orbital evolution.

However, the dissipation of the $0.6~\Msun$ star is initially one order of magnitude smaller than the two other stars.
After a few million years and for the rest of the evolution, the dissipation of the $0.6~\Msun$ becomes more than one order of magnitude higher than the two other stars. 
However, the tidal evolution timescale does not only depend on the dissipation of the star but also on its radius. 
Equation \ref{Tp} shows that the smaller the radius of the star, the longer the evolution timescales.
Thus, despite a higher dissipation, the tidal evolution around the $0.6~\Msun$ star occurs on longer timescales than the more massive stars.

\smallskip

For high mass planets, such as the Jupiter-mass planet of Fig. \ref{comp_dissip_sma_spins_sigmasbar_Mp_2_318_Ms_smooth_1d8}b, the migration occurs on shorter timescales as discussed in the previous section. 
The difference between a star of $0.6~\Msun$ and the two more massive ones is more pronounced.
The planet orbiting the lowest mass star migrates outward less than for the other stars but does so for a longer time.
This is due to the smaller radius of the $0.6~\Msun$ star which does more than compensate the higher dissipation.
Due to the increase of the dissipation happening in the first few $10^7$~yr of evolution, the migration of the planet first accelerates.
Then as the dissipation $\oss$ stabilizes, the migration decelerates due to the increase of the semi-major axis and the consequent increase of the evolution timescale.

The evolution of the planet orbiting the $1~\Msun$ star is slightly different than the one orbiting the $1.2~\Msun$ star.
Migration initially happens faster around a $1.2~\Msun$ star than a $1~\Msun$ star because the radius of the higher mass star is bigger (their dissipation being of the same order of magnitude). 
However as time passes, the evolution timescale of the planet orbiting the $1.2~\Msun$ star increases first because the planet migrates outward and second because the dissipation $\oss$ decreases by almost two orders of magnitude.
In the end, the outward migration stops around an age of 30~Myr at a semi-major axis of $\sim0.05$~AU.
For the $1~\Msun$ star, the combination of a smaller radius and a higher dissipation leads to a slower evolution on a longer time so that the planet reaches $\sim 0.05$~AU around an age of $\sim 200$~Myr.

 
\section{Rotational evolution of the host star}\label{star_rot_evol}

During their orbital evolution, the planets influence the rotation of their host star. 
This influence is more or less visible depending on the initial semi-major axis and the mass of the planet. 
One extreme exemple can be seen in Figure \ref{comp_dissip_sma_spins_sigmasbar_Mp_10_Ms_1_smooth_1d8}b, when the evolution is driven by the equilibrium tide. 
The close-in planets collide with the star in a few gigayears and this entails an acceleration of the rotation of the star \citep[see][for a description of this phenomenon]{Bolmont2012}.
However, when the evolution is calculated with our model and the star is initially rotating fast, the influence is less visible as it leads to much less collisions.

\subsection{Influence of the planet on the rotation of the star}

We investigated the influence of the planet on the rotation of the star for each star considered here.
In order to quantify the effect of the planet on the spin of the star, we introduce the quantity $\delta P$ defined as follows:
\begin{equation}
\delta P = P_{\star, {\rm tides = on,~t = }5~{\rm Gyr}} - P_{\star, {\rm tides = off,~t = }5~{\rm Gyr}},
\end{equation}
where $P_{\star, {\rm tides = on,~t = }5~{\rm Gyr}}$ is the rotation period of the star at an age of $5~$Gyr for simulations with our tidal model and $P_{\star, {\rm tides = off,~t = }5~{\rm Gyr}}$ is rotation period of the star at an age of $5~$Gyr for simulations with only the influence of the wind-braking.

\begin{figure}[htbp!]
        \begin{center}
        \includegraphics[width=9cm]{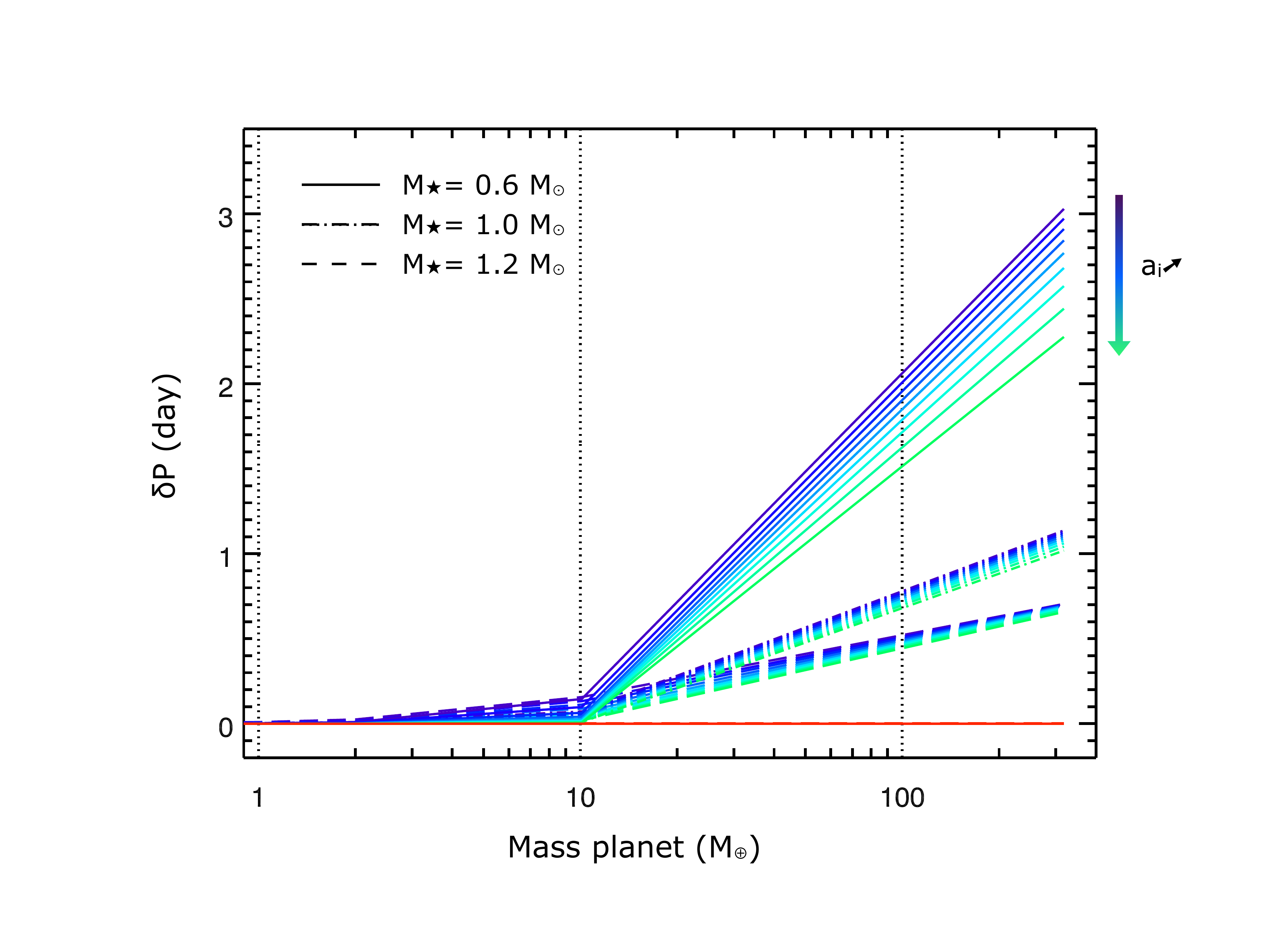}
        \caption{$\delta P$ for different initial orbital distances and for the 3 different stars. The full lines correspond to the $0.6~\Msun$ star, the dashed-dotted lines correspond to $1.0~\Msun$ and the long-dashed line correspond to $1.2~\Msun$. For purple to green: an initial semi-major axis of $0.024$, $0.026$, $0.028$, ..., $0.040$~AU. The red line marks $\delta P = 0$~day. These simulations were done with a regularization value $\rho$ of $10^{-5}$~s$^{-1}$.}
        \label{comp_dissip_spinsf_vs_Mp_smooth_1d5_article}
        \end{center}
        \end{figure}

Figure \ref{comp_dissip_spinsf_vs_Mp_smooth_1d5_article} shows the variation of this quantity with the mass of the planet, for different initial semi-major axes: $0.024$, $0.026$, $0.028$, ..., $0.040$~AU and for the 3 different stars considered here.
{Only the spin of the stars whose planets have survived the 5~Gyr evolution is plotted.
As the star considered here is an initially fast rotator, the planets considered here all migrate outwards during their evolution.
We discuss in the following section what happens when the planets fall towards the star, which occurs when the star is an initially slow rotator.}
We find that the more massive the star, the less influence the planet has on the stellar spin evolution; and the more massive the planet, the more influence the planet has on the stellar spin evolution. 
We retrieve here the signature of the applied torque, which scales for a circular orbit as $(\Mp/\Ms)^2 \Ms \Rs^2\left(\Rs/a\right)^6\mathbf{(\Os-n)}$ \citep[][]{Zahn1989}. 
The planets have here the effect of slowing down the rotation of the star. 
For planets less massive than $10~\Mearth$, the effect of the planet on the spin of the star is negligible.
For $\Mp = 10~\Mearth$, the difference at 5~Gyr is of a few hours, while for $\Mp = 1~\Mjup = 317.8~\Mearth$ it can reach several days.
A hot Jupiter around an initially fast rotating Sun-like star can slow down the rotation of its star by a day.

We can see that $\delta P$ is bigger for a Jupiter-mass planet orbiting a $0.6~\Msun$ star, where the difference can be up to 3 days for the close-in planets.
With such a difference, the influence a planet has on the stellar spin can be potentially measured. 
Indeed, high resolution photometric space missions such as CoRoT \citep{Baglin2006} and {\it Kepler} \citep{Borucki2010} allows us to precisely measure stellar rotation \citep[a sensibility up to $10\%$ of the rotation period; e.g.][]{Garcia2014} and to test proposed gyrochronology relationships \citep[e.g.][]{Gizonetal2013, McQuillan2013, Paz2015, Ceillieretal2016, VanSadersetal2016}. 
For example, \cite{Ceillieretal2016} demonstrated that the rotation of KOIs hosting low-mass planets are the same, up to the observational sensitivity, that assumed single similar solar-type stars. This result is coherent with our theoretical prediction.

\subsection{Influence of initial rotation spin}\label{init_spin}

\begin{figure*}[htbp!]
	\centering
        \includegraphics[width=17cm]{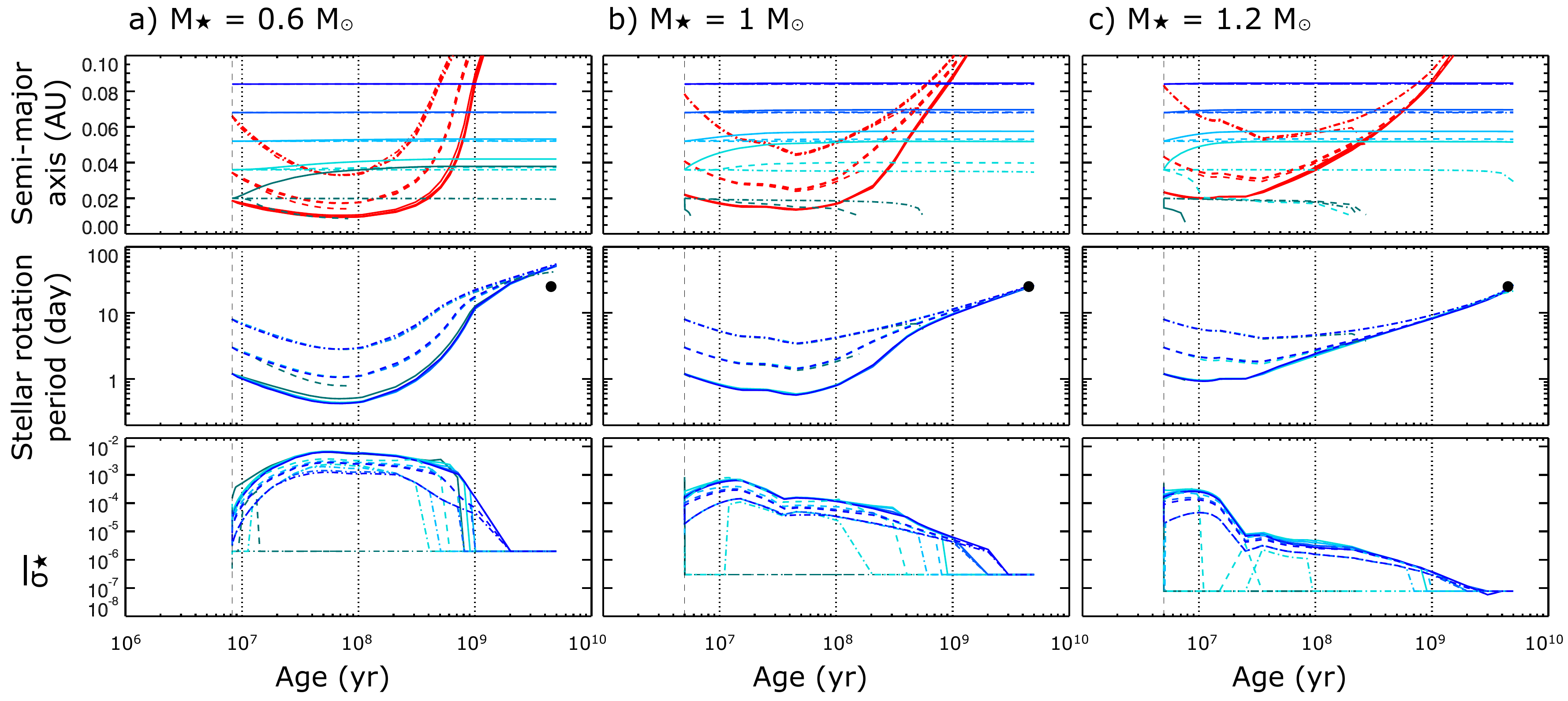}
        \caption{Tidal evolution of a $\Mjup$ planet around a a) $0.6~\Msun$, b) $1~\Msun$, c) $1.2~\Msun$ star for different stellar initial rotation periods: 1.2~day (full lines), 3~day (dashed lines) and 8~day (dashed-dotted lines). The initial time for each of the simulations is represented in thin vertical black dashed lines. Top: evolution of the semi-major axis of the planets (green to blue lines), corresponding corotation radius (red lines). Middle: rotation period of the star, the rotation of the Sun at present time is represented with a black dot. Bottom: Evolution of the normalized dissipation factor $\oss$ (full lines) and the equilibrium tide dissipation factor (dashed lines). These simulations were done with a regularization value $\rho$ of $10^{-5}$~s$^{-1}$.}
        \label{comp_dissip_sma_spins_sigmasbar_Mp_318_Ms_Ps0_smooth_1d5}
        \end{figure*}

\citet{Bouvier1997} were considering two populations of stars which are thought to bracket to the possible state of stars just after the disk dissipation: the initially fast rotators ($P_{\star, 0}$ = 1.2~day) and the initially slow rotators ($P_{\star, 0}$ = 8~day).
Following \citet{Bolmont2012}, we investigated here the influence of the initial rotation spin of the star on the outcome of the planets considering an initially fast rotator ($P_{\star, 0}$ = 1.2~day), an initially slow rotator ($P_{\star, 0}$ = 8~day) and a star with an intermediate initial spin ($P_{\star, 0}$ = 3~day).
Figure \ref{comp_dissip_sma_spins_sigmasbar_Mp_318_Ms_Ps0_smooth_1d5} shows the evolution of a Jupiter-mass planet orbiting the three different stars and for the three different initial stellar rotation periods.
The planets initially around faster rotating stars migrate farther away and the more massive the star, the bigger are the differences between the different stellar initial spins.

For $0.6~\Msun$, the planets initially closer than $0.05$~AU undergo a different evolution depending on the initial rotation state of the star.
It is especially true for the innermost planet: it survives and migrates outward for $P_{\star, 0} = 1.2$~day, it falls onto the star for $P_{\star, 0} = 3$~day and it does not experience significant orbital migration for $P_{\star, 0} = 8$~day.
For $P_{\star, 0} = 1.2$~day, it is initially exterior to the corotation distance and in the region where $P_{\rm orb} > 1/2 P_\star$ so it rapidly migrates outward due to the dynamical tide. 
However, for $P_{\star, 0} = 3$~day, it is initially interior to the corotation distance so it migrates inward and as it is also in the region where $P_{\rm orb} < 1/2 P_\star$, the equilibrium tide is initially driving the evolution.
However as the star spins up, the planet reaches the limit $P_{\rm orb} = 1/2 P_\star$ and the dynamical tide takes over. 
The planet stays for a few 10 million years on the limit $P_{\rm orb} = 1/2 P_\star$ following the mechanism introduced in Section \ref{evol_1Msun} before falling onto the star.
Note that while falling onto the star, the planet makes the star spin up, which is visible on the second panel of Figure \ref{comp_dissip_sma_spins_sigmasbar_Mp_318_Ms_Ps0_smooth_1d5}c with the dark green dashed line.
Finally for $P_{\star, 0} = 8$~day, the planet is initially interior to the corotation radius so it migrates inward, however contrary to the previous case the star does not spin up enough, so that the condition $P_{\rm orb} = 1/2 P_\star$ is never met. 
The planet therefore evolves solely due to the equilibrium tide, and as the equilibrium tide dissipation is small, the evolution timescales are very long and the planet essentially remains at its initial semi-major axis. 
For planets farther away, the only difference between the different stellar initial spins is the final semi-major axis. 
This qualitative difference vanishes for planets initially at more than $0.05$~AU. 

When the planet evolves due to the dynamical tide, the dissipation $\oss$ follows the same kind of evolution independently of the initial stellar rotation.
However there is a quantitative difference due to the difference of spin (Eq. \ref{delta_tau}), which entails an order of magnitude of difference for $\oss$ between the fast rotator and the slow rotator.
Planets orbiting the slow rotating star ($P_{\star, 0} = 8$~day) leave the region $P_{\rm orb} > 1/2 P_\star$ earlier in the evolution of the star than for the other stars. 
In the end, the evolution around slowly rotating stars is slower for two reasons: the slower rotation and the fact that the dynamical tide drives the evolution for a shorter time.
 
 \smallskip
 
The higher the mass of the star, the higher the corotation radius for the same initial rotation period. 
Consequently the close-in planet (at $0.02$~AU), which was just outside the corotation radius when orbiting the $0.6~\Msun$ star initially rotating at $P_{\star, 0} = 1.2$~day, is just inside the corotation radius when orbiting the $1.0~\Msun$ star initially rotating at $P_{\star, 0} = 1.2$~day.
Consequently, the planet falls onto the star.
And it does so independently of the initial stellar spin. 
The only difference is that the faster the star is rotating, the quickest the planet falls: about 1~Myr for $P_{\star, 0} = 1.2$~day, a few 10~Myr for $P_{\star, 0} = 3$~day and a few 100~Myr for $P_{\star, 0} = 8$~day.
For farther out planets, the only difference is the final semi-major axis, which is clearly visible for the planet beginning its evolution at $0.036$~AU. 
Indeed, this planet migrates slightly inward for $P_{\star, 0} = 8$~day, it migrates outward up to $0.04$~AU for $P_{\star, 0} = 3$~day, and up to $0.05$~AU for $P_{\star, 0} = 1.2$~day.
Planets initially farther than $0.07$~AU undergo the same evolution (= no orbital evolution) independently of the initial rotation state of the star.
For a star of $1.2~\Msun$, the differences with $0.6~\Msun$ are even more marked.
Due to the higher mass and consequent increase of the corotation radius the planet initially at $0.036$~AU experiences different evolution depending on the initial spin: it survives and migrates away when $P_{\star, 0} = 1.2$~day, it falls onto the star when $P_{\star, 0} = 3$~day and migrates in on a gigayear timescale when $P_{\star, 0} = 8$~day.

\smallskip

Figure \ref{comp_dissip_sma_spins_sigmasbar_Mp_318_Ms_Ps0_smooth_1d5} also shows that hot Jupiters can be found at small orbital distances around initially slow rotating stars.
If formation mechanisms in the protoplanetary disk allow to form hot Jupiters equally around all three type of stars, we therefore find that they are more likely to survive on gigayear timescales around low-mass stars rather than higher mass stars.
This statement is in contradiction with the observations that show that the occurrence rate of hot Jupiters around low mass stars is very low \citep[e.g.][]{Bonfils2013}.
The difference in the occurrence rate of hot Jupiters around low mass stars compared to Sun-like stars must therefore depend on tidal dissipation in the stellar radiative core \citep{Guillotetal2014} or on formation mechanisms rather than on the following tidal evolution related to the dissipation in the convective enveloppe.
For an initially slowly rotating $1~\Msun$ star, Jupiter mass planets survive a 5~Gyr evolution if they start their evolution at a semi-major axis bigger than $0.028$~AU. 
However these planets would be currently migrating towards the star. 
For exemple, the orbit of a hot Jupiter beginning at $0.028$~AU around an initially slowly rotating star shrinks to $0.024$~AU after $\sim 5.3$~Gyr of evolution.

        \begin{figure}[htbp!]
        \begin{center}
        \includegraphics[width=9cm]{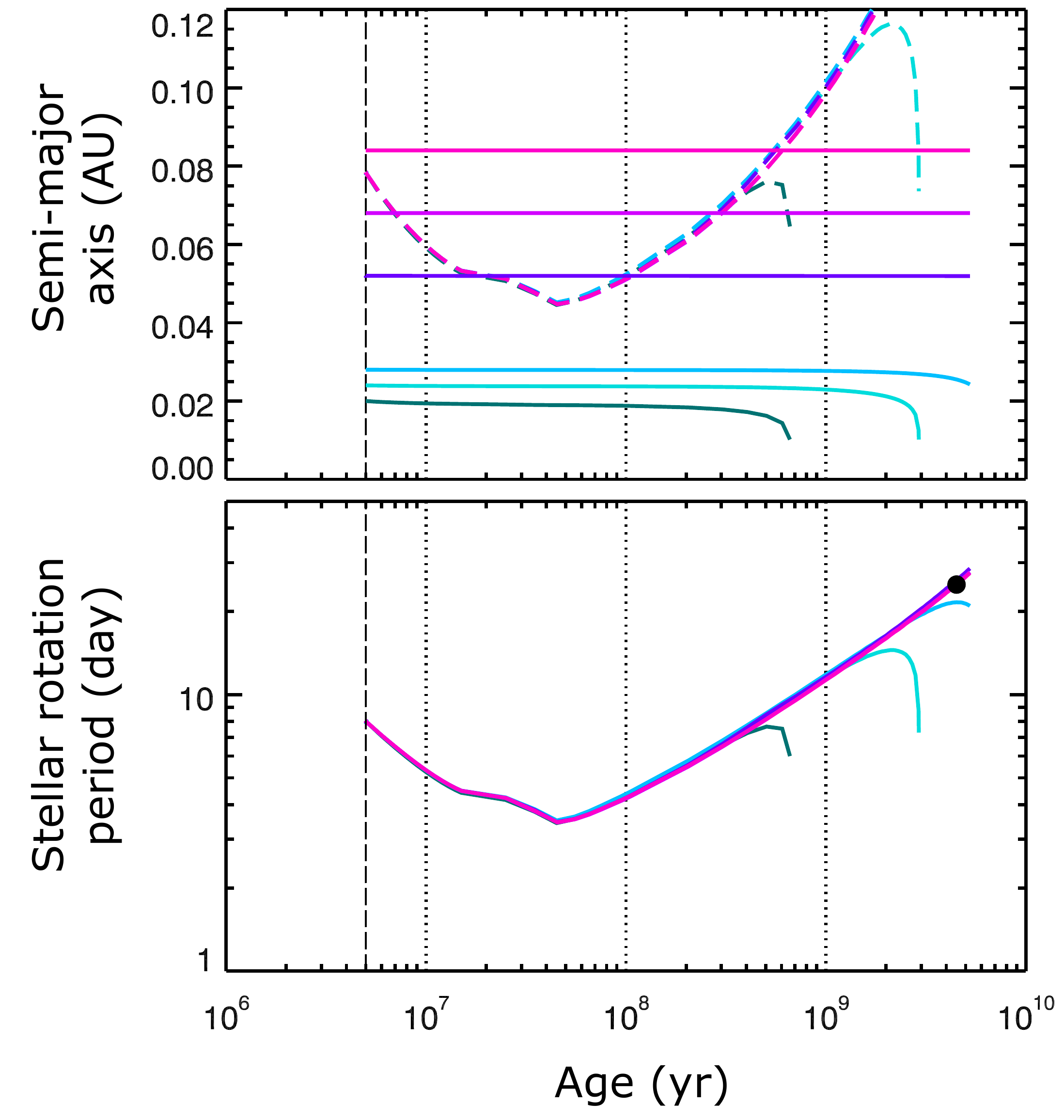}
        \caption{Tidal evolution of a $1~\Mjup$ planet around an initially slow rotating $1~\Msun$ star. The initial time for the simulations is 5~Myr. Top: evolution of the semi-major axis of the planets (full lines), corresponding corotation radius (dashed lines). Bottom: rotation period of the star, the rotation of the Sun at present time is represented with a black dot. These simulations were done with a regularization value $\rho$ of $10^{-5}$~s$^{-1}$.}
        \label{comp_dissip_tinit_sma_spins_sigmabar_Mp_318_Ms_1_smooth_1d5_fall_or_not}
        \end{center}
        \end{figure}

\medskip

{Due to tidal interaction, the massive planets either fall or migrate away. 
When falling onto the star, the planets make their host star spin-up.
This phenomenon was discussed in \citet{Pont2009} and \citet{Bolmont2012}.
Figure \ref{comp_dissip_tinit_sma_spins_sigmabar_Mp_318_Ms_1_smooth_1d5_fall_or_not} shows the evolution of the semi-major axis of planets orbiting a $1~\Msun$ star with an initial rotation of 8~day, as well as the evolution of the corresponding stellar spin. 
The planets can be sorted into two populations: three planets are falling onto the star and three planets survive the 5~Gyr evolution.
Among the three falling planets, two of them fall onto the star in less than 2~Gyr, making the star spin up significantly. 
For example, the planet falling when the star is 2~Gyr makes the star spin up to $\sim7$~day, when it should have had a rotation of $\sim20~$day.
The planet which is still falling at an age of 5~Gyr is responsible for an spin-up of about 8~day with respect to the ``normal'' rotation period at that moment.
This difference will increase as the planet continues falling towards the star.
} 

The ingestion of planets by the star has been one explanation proposed for the observed dearth of close-in planets around fast rotating stars (see \citealt{McQuillan2013} for the observational aspect and \citealt{Teitler2014} for a theoretical explanation using a standard equilibrium tide model).
An alternative explanation based on a combination of secular orbital evolution in a multiple planet system and tides was suggested by \citet{LanzaShkolnik2014}.
They showed that around old stars -- and thus according to gyrochronology slowly rotating stars -- planets initially far away and excited on a very eccentric orbit could have had time be tidally circularized on short orbits.
We show here that due to the stronger tidal dissipation induced by dynamical tide, this observed dearth can also be partly accounted by the outward migration of the planets which depletes the inner parts of a system. 
We will investigate in the future the effect of this higher dissipation on the evolution of multiple planet systems using the \textit{Mercury-T} code \citep{Bolmont2015}.




\section{Conclusions}
\label{Discussion}

We present here a new tidal model, which constitutes a first step in the endeavor to reconcile two tidal formalisms.
This improvement constitutes a first step because 1) it is using an averaged model for the quality factor of the star \citep[see][]{Mathis2015b} and 2) it is only valid for non-inclined circular systems.
This new model takes into account the effect of the dynamical tide, and more specifically the evolution of the dynamical tide-induced dissipation in the convective enveloppe over the evolution of the host star.\\

We found that due to the consequent enhanced dissipation compared to a standard equilibrium tide dissipation, the planets experience a more pronounced tidal orbital evolution.
For example, a Jupiter mass planet orbiting a fast rotating star can migrate from $\sim 0.02$~AU to $0.05$~AU when its evolution is due to the dynamical tide instead of falling onto the star when its evolution is due to the equilibrium tide.
This changes the conclusions of the work of \citet{Bolmont2012} and underlines the need to incorporate better tidal models in orbital dynamics codes.

We found that the planets have behaviors more similar to what was first introduced in \citet{Bolmont2011} than to what was shown in \citet{Bolmont2012}.
This means that the dynamical tide is strong enough so that planets can be influenced by the early age spin evolution of the star.
Indeed, a planet with an initial orbital distance much smaller than the corotation radius will migrate inwards to eventually fall onto the star.
A planet initially exterior to the corotation radius will migrate outward. 
Finally, a planet initially interior to the corotation radius but sufficiently close to it can be saved by the spin-up of the star during the PMS phase (and the consequent shrinking of the corotation radius).  

We also found that planets initially interior to the corotation radius fall onto the star in three steps.
First, the dynamical tide brings the planet to the limit $P_{\rm orb} = 1/2 P_\star$. 
Second, a competition between the acceleration of the rotation of the star and the equilibrium tide-induced inward migration causes the planet to remain at $P_{\rm orb} = 1/2 P_\star$.
Third, when the planet has sufficiently migrated inward, the equilibrium tide is strong enough to make the planet fall onto the star.

We investigated the influence of the initial spin of the star on the orbital evolution of the planets and we found that for a Jupiter-mass planet the tidal evolution around a star initially rotating at $P_{\star, 0} = 1.2~$day, 3~day and 8~day is different if the planet is closer than $\sim0.07$~AU. 
Depending on the initial spin, a close-in planet can either fall onto the star, migrate outward or not experience any significant orbital evolution.
For a planet of $1\Mearth$, this limit decreased to $\sim0.03$~AU.
We found that ``really'' hot Jupiters ($P_{\rm orb}<5~$day) would be more common around initially slow rotating stars, and if their formation rate was independent on the stellar mass, they would be more common around low mass stars.
This illustrates the fact that the scarcity of hot Jupiters around low mass stars may be due to the dissipation in the radiative core or to formation mechanisms within the protoplanetary disk.

We found that the planet has a negligible effect on the spin rotation history of the star unless its mass is higher than $10~\Mearth$.
For a $1~\Msun$ star, the presence of a hot Jupiter {around an initially fast rotating star} can lead to a spin-down of the star of one day at an age of 5~Gyr. 
For an initially fast rotating $0.6~\Msun$ star, the spin-down can be up to three days. 
{However, when considering initially slow rotating stars, more planets are engulfed causing a spin-up of the stellar rotation.
Consequently, observations of the spin of stars hosting hot Jupiters for which we know the age (through another mean than gyrochronology) could eventually tell us something about the rotation history of the star: if the star is rotating \textit{slower} than it would without planets, it means that it was probably initially rotating \textit{fast}; if however the star is rotating \textit{faster} than it would without planets, it means that it was probably initially rotating \textit{slow}.}
This can be now probed thanks to high-precision asteroseismology. 
Note that the modification of the rotational evolution of a low-mass star due to the presence of a massive companion may also have several important consequences. First, because of the applied tidal torque, the angular momentum redistribution inside the star may be modified in comparison to the case of a single star \citep[see e.g.][]{Zahn1992,Zahn1994,MZ2004,MR2013,Amardetal2016}. Then, internal differential rotation profile may be different, leading to a different mixing of chemical elements and thus different surface chemical abundances at the surface \citep[][]{Zahn1994}. Next, it may affect the magnetic activity of cool and solar-type stars. Indeed, surface magnetic fields observed at the surface of low-mass stars are generated by a dynamo action occurring in their external convective zone, the amplitude, the geometry and the possible cyclic behavior of the fields being strongly correlated with the mass, the age and the rotation of stars \citep[see e.g.][]{Charbonneau2014,Brun2014}.\\

As explained before, this model is a first step towards a coherent description of tidal dissipation in orbital dynamics codes. 
First, as discussed in \S \ref{tidal_model_1}, dissipation of tidal waves in the radiative core of low-mass stars hosting planets as well as the angular momentum exchanges with the surrounding convective enveloppe should be taken into account. 
Next, additional complex physical mechanisms that may impact tidal flows and their dissipation such as differential rotation \citep[][]{BaruteauRieutord2013, Favieretal2014, Gueneletal2016} and magnetic fields \citep[][]{Barkeretal2014} have to be studied. The same effort should also be undertaken for planetary interiors to have tidal models taking into account their structural and dynamical properties \citep[e.g.][]{Correiaetal2003, OgilvieLin2004, Tobieetal2005, Henningetal2009, Efroimsky2012, RMZL2012, Correiaetal2014, Gueneletal2014}. 
The second important step, will be to take into account dependence of tidal dissipation/torques on tidal frequency. 
It will allows us to treat the case of eccentric orbits and avoid any singularity of tidal models. 
Finally, it would be important to take into account orbital inclination and planetary obliquity \cite[e.g.][]{Lai2012}.

Furthermore, the wind prescription we used in this work \citep{Bouvier1997} can also be improved for example by using a more recent parametrization which depends closely on the structural evolution of the star \citep{Matt2015} and the geometry of the field \citep{Reville2015}.
This would be particularly useful in order to have a more precise insight of the influence of the planets on the spin history of the star. 
Finally, we will use our model to try to reproduce the orbital and stellar spin state of known systems \citep[as was done in][which found similar results as \citealt{Bolmont2012} and in \citealt{DamianiLanza2015}]{Ferraz-Mello2015}.


\begin{acknowledgements}  
We thank the referee for the useful comments.
E. B. acknowledges that this work is part of the F.R.S.-FNRS ``ExtraOrDynHa'' research project. S. M. acknowledges funding by the European Research Council through ERC grant SPIRE 647383. This work was also supported by the ANR Blanc TOUPIES SIMI5-6 020 01, the Programme National de Plan\'etologie (CNRS/INSU) and PLATO CNES grant at Service d'Astrophysique (CEA-Saclay).
\end{acknowledgements}


\end{document}